\documentstyle[11pt,aaspp4]{article} 
\lefthead{Hubeny \& Hubeny}
\righthead{Non-LTE models of AGN disks}

\begin{document}

\title{Non-LTE Models and Theoretical Spectra of Accretion Disks in
Active Galactic Nuclei}

\author{ Ivan Hubeny\altaffilmark{1} and Veronika Hubeny\altaffilmark{2}}
\affil{Laboratory for Astronomy and Solar Physics,
       NASA Goddard Space Flight Center, Greenbelt, MD 20771 \\
       E-mail: hubeny@stars.gsfc.nasa.gov, 
       veronika@physics.ucsb.edu}

\altaffiltext{1}{AURA/NOAO.}
\altaffiltext{2}{Present address: Department of Physics, 
University of California, Santa Barbara, CA 93106}

\authoraddr{NASA/GSFC, Code 681, Greenbelt, MD 20771}

\begin{abstract}

We present self-consistent models of the vertical structure and emergent
spectrum of AGN accretion disks. The central object is assumed to be
a supermassive Kerr black hole. 
We demonstrate that NLTE effects and the effects of a self-consistent
vertical structure of a disk play a very important role in determining
the emergent radiation, and therefore should be taken into account. 
In particular, NLTE models exhibit a largely diminished H I Lyman 
discontinuity when compared to LTE models, and 
the He II discontinuity appears strongly in emission for NLTE models.
Consequently, the number of ionizing photons in the He~II Lyman
continuum predicted by NLTE disk models is by 1 - 2 orders of magnitude 
higher than that following from the black-body approximation.
This prediction has important implications for ionization models of AGN
broad line regions, and for models of the intergalactic radiation 
field and the ionization of helium in the intergalactic medium. 

%
% ====================================================
%

\end{abstract}


\clearpage

%
% ====================================================
%
%\twocolumn

\section{INTRODUCTION}

Accretion disks around massive black holes are believed to provide the
ultraviolet and soft X-ray flux observed in many Active Galactic Nuclei (AGN).
The observational evidence 
is based on the `big blue bumps' seen in the ultraviolet (e.g. Shields 1978; 
Malkan \& Sargent 1982). Other observations, however, indicate that the
H I Lyman jump is weak or non-existent (e.g. Antonucci, Kinney, \& Ford 1989),
which has been used to argue against the presence of a
geometrically thin, optically thick accretion disk around a black hole.

To settle the argument, one has to construct realistic models
of accretion disks and their emergent spectra, and to check 
whether the predicted spectrum is consistent with observations.
Several models with varying degrees of sophistication were presented in the 
past.
The simplest approach for calculating an accretion disk spectrum is to assume
that each point of the disk radiates as a blackbody at the local effective
temperature. The blackbody approach is
acceptable for studying the basic energetics of the system, but is obviously 
inadequate for predicting individual spectral
features. To improve the situation, Kolykhalov \& Sunyaev (1984), Sun (1987),
and  Sun \& Malkan (1989) have used model stellar atmospheres 
to describe the radiation from different parts of an AGN disk. 
This would offer a very attractive method of computing the AGN spectra, since 
a large pool of model stellar atmospheres exists. However, one should be
cautious about using this approach because the structure of a disk and
and a stellar atmosphere may be significantly different.

To explore this problem, 
several authors (e.g. Laor \& Netzer, 1989; Ross, Fabian, \& Mineshige 
1992; St\"orzer and Hauschildt 1994; Coleman 1994; Shields \& Coleman 1994;
Blaes \& Agol 1996) have constructed models of the vertical structure
of AGN accretion disks, using however various simplifying approximations.
The most realistic models computed so far are those of St\"orzer, Hauschildt
\& Allard (1994), and by D\"orrer et al. (1996). 
The former authors (and also Coleman 1994 and Shields \& Coleman 1994)
specifically addressed the question of how departures from Local 
Thermodynamic Equilibrium (LTE) influence the predicted H I Lyman
discontinuity, and showed that non-LTE effects reduce the strength of the H I
Lyman jump. 
However, they used an approximate, semi-analytical approach
to solve for the physical structure of the disk. D\"orrer et al. have
solved the vertical structure self-consistently, using however two
critical simplifications, i) LTE, and ii) no bound-free opacities, and a  
pure-hydrogen atmosphere. On the other hand, they treat the Compton
scattering in detail.

In this {\it Letter}, we present some representative self-consistent, 
non-LTE models of the vertical structure of AGN accretion disks. 
The basic aim of this
study is to investigate differences in the predicted spectrum between this and
simpler approaches, and to show that simplified models may lead to inaccurate
and misleading conclusions. 
In order to emphasize the observable
consequences of self-consistent, non-LTE models, we present the predicted
spectra for a few representative rings in the disk. 
Integrated spectra of the whole disk,
taking into account general relativistic photon transfer functions
(Cunningham 1975; Speith, Riffert, \& Ruder 1995), will be considered 
in a separate paper.

%
% ====================================================
%

\section{THE MODEL}

We assume a steady-state, geometrically thin disk in a (general relativistic)
``Keplerian'' rotation. 
The disk is divided into a set of axially symmetric concentric rings, 
each ring behaving as a 1-D radiating slab.
The relativistic radial disk structure was calculated by Novikov \& Thorne
(1973), Page \& Thorne (1974), and recently by Riffert \& Herold (1995),
who have corrected the previously derived form of the vertical pressure 
balance equation (the correction was first worked out by Eardley \& Lightman
1975).
We will use the results of Riffert \& Herold (1995).

The vertical structure of a single ring is computed by solving simultaneously
the hydrostatic equilibrium equation, the energy balance equation, the
radiative transfer equation, and, since we do not assume LTE, 
the set of statistical equilibrium equations. 

If we neglect self-gravity of
the disk, and assume that the radial distance from the black hole, $R$,
is much larger than the distance from the central plane, $z$, 
we can write the equation of hydrostatic
equilibrium as
\begin{equation}
\label{he}
{dP \over dz} = -\rho g\enspace ; \qquad 
g = {GM \over R^3 }{C \over B} \, z \, ,
\end{equation}
where $P$ is the total pressure, $\rho$ is the mass density, $M$ is the 
black hole mass, and $G$ is the gravitational constant. $B$ and $C$ 
(and also $A$ and $D$ used later) are
relativistic corrections in the notation of D\"orrer et al. (1996).
The total pressure is given as a sum of the gas pressure and the radiation 
pressure,
\begin{equation}
\label{press}
P=P_{\rm gas} + P_{\rm rad} = NkT + 
{4\pi\over c} \int_0^\infty K_\nu \, d\nu \, ,
\end{equation}
where $N$ is the total particle number density, $T$ the temperature,
$k$ the Boltzmann constant, and $K_\nu$ the second moment of the specific
intensity of radiation $I_\nu(\mu)$, i.e. 
$K_\nu = \int_{-1}^1 \mu^2 I_\nu(\mu) d\mu/2$;
$\mu$ being the cosine of the angle between direction of propagation and
the normal to the disk midplane.
The upper boundary condition is taken from Hubeny (1990a -- Eqs. 4.19-4.20 
there).

The energy equation expresses the balance between the mechanical energy
dissipated by viscous shearing between the ``Keplerian'' orbits and the net
radiation loss, viz.
\begin{equation}
\label{ener}
{9 \over 4} {G M \over R^3} \left( {A \over B} \right)^2 \, \rho \,w
=4 \pi \int_0^\infty (\eta_\nu - \kappa_\nu J_\nu) d\nu  \, ,
\end{equation}
where $w$ is the viscosity, and $\eta_\nu$ and $\kappa_\nu$
are the monochromatic absorption and emission coefficients; 
$J_\nu $ is the mean intensity of radiation. 
We parameterize the viscosity using a fixed Reynolds number
(Lynden-Bell \& Pringle 1974; Hubeny 1990a). 
We will show in a subsequent paper that this
treatment yields very similar results to the traditional
$\alpha$--parametrization of viscosity, introduced
by Shakura \& Sunyaev (1973), and recently modified for AGN disks by
D\"orrer et al. (1996).
The total energy flux from the disk surface is given through the
effective temperature,
\begin{equation}
\label{teff}
T_{\rm eff} = \left( {3 G M \dot M \over 8 \sigma \pi R^3}{D \over B} 
\right)^{1/4}\, ,
\end{equation}
where $\dot M$ is the mass accretion rate, and $\sigma$ the
Stefan-Boltzmann constant.

The radiative transfer equation is written in the standard way (e.g. Mihalas
1978), viz
\begin{equation}
\label{rte}
\mu {d I_\nu(\mu) \over d\tau_\nu} = I_\nu - S_\nu  \, ,
\end{equation}
where the monochromatic optical depth is defined through 
$d \tau_\nu \equiv -\chi_\nu dz$; 
$S_\nu \equiv \eta_\nu/\chi_\nu$ 
is the source function. 
Here, $\chi_\nu$ is the total
absorption coefficient, $\chi_\nu= \kappa_\nu+\sigma_\nu$, $\sigma_\nu$
being the scattering coefficient. We assume that the only
scattering process is the electron (Thomson) scattering. Finally, the emission
coefficient is given by
$\eta_\nu = \eta_\nu^{\rm th} + \sigma_\nu J_\nu$,
where $\eta_\nu^{\rm th}$ is the coefficient of thermal emission.
We assume no incident radiation at the disk surface, and a symmetry condition
at the disk midplane, $I_\nu(\mu)=I_\nu(-\mu)$.

In LTE, the {\em thermal} component of the source function 
is given by  $S^{\rm th} \equiv \eta_\nu^{\rm th}/\kappa_\nu = B_\nu$, 
with $B_\nu$ being the Planck function, 
i.e. it is a simple function of the local temperature. 
However, it is well known that the LTE approximation breaks down in low-density,
radiation-dominated media, which are precisely the
conditions prevailing in the AGN disks. Therefore, we have to adopt a more
general treatment, traditionally called non-LTE (or NLTE), where 
the thermal source function deviates from the Planck function, because 
the populations of energy levels are allowed to
depart from the Boltzmann-Saha distribution. These populations are determined
through the equations of statistical equilibrium.

The overall system of structural equations
forms a highly coupled, non-linear set of 
integro-differential equations. These are, however, very similar
to the equations describing a classical NLTE stellar atmosphere
(e.g., Hubeny 1990a,b). We may therefore adopt numerical methods
and computer programs developed for stellar atmospheres.
We use here the computer program TLUSDISK, which is a derivative of the stellar
atmosphere program TLUSTY (Hubeny 1988; Hubeny \& Lanz 1995).
We stress that no a priori assumptions about the height of the disk 
and its total optical thickness are made; these are determined
self-consistently with other structural parameters.

A NLTE model of the vertical structure of one ring of a disk is 
computed in three
steps. First, an LTE-gray model is constructed, as described in Hubeny (1990a).
This serves as the starting solution for the subsequent step, 
an LTE model, computed by TLUSDISK. This in turn is used as the starting
solution for the last step, a NLTE model.
In the NLTE step, we assume that all bound-bound transitions are in 
detailed radiative balance. To stress this point, we denote these models as
NLTE/C (i.e. NLTE with continua only).
Since our primary interest here is to explore the continuum flux around the
H I and He II Lyman discontinuities, 
which are formed at layers where the lines are indeed in detailed balance,
this approximation is justified.

We consider here disks composed of H, He, C, N, and O,
with solar abundances.
The model atoms are the same as in Hubeny \& Lanz (1995), with 9 NLTE
levels of H, 1 level of H${}^+$, and 14, 14, and 1 levels for He, He${}^+$,
and He${}^{++}$, respectively, and with a total of about 50 levels for
C III - C V, N III - N V, and O III - O VI. 
The opacity sources we consider are the
bound-free transitions from all considered levels of H, He, C, N, O; 
the free-free opacity of the considered ions, and the electron (Thomson) 
scattering. 
All the relevant cross-sections and relevant collisional rates are summarized
in Hubeny (1988). 
In this paper, we do not consider the Compton scattering because it is only
marginally important for the considered models; 
it will however be considered in the subsequent paper.

%
% ====================================================
%

\section{RESULTS}

As a representative case, we take a disk around a Kerr supermassive black hole
with $M = 2 \times 10^9 M_{\odot}$, with a limiting stable rotation
(the specific angular momentum $a/M=0.998$ -- Thorne 1974).
The accretion rate is taken to be $\dot{M}=1~ M_\odot{\rm yr}^{-1}$.
We have calculated a number of vertical structure models for rings at
various distances; we present here two models for
$R/R_{\rm g} = 2$ and 4, where $R_{\rm g}$ is the gravitational radius,
$R_{\rm g} = G M/c^2$. These rings 
are representative for providing the emergent soft X-ray and EUV radiation
of the disk.
Complete results will be presented elsewhere (Hubeny \& Hubeny, in prep.).

Our basic aim is to explore the effects of NLTE on the emergent spectra.
First, we will compare the spectra computed for self-consistent models of
the vertical structure of the disk as calculated in LTE and NLTE.
We will also present theoretical spectra for a ``partial non-LTE'' model,
i.e. the one with the vertical structure (temperature and density) fixed by
an LTE model, and where the NLTE effects are included only in solving
simultaneously the radiative transfer and statistical equilibrium equations
(e.g., Coleman 1994; Shields \& Coleman 1994; 
St\"orzer, Hauschildt, \& Allard 1994).
Next, we will present an emergent radiation
from a classical NLTE (plane-parallel, hydrostatic)
stellar atmosphere model, computed for the same effective
temperature as the disk ring. The surface gravity should be taken to be the 
one corresponding
to the local disk gravity at some characteristic layer (e.g. where the 
Rosseland optical depth is unity).  
However, such an atmosphere is usually 
unstable (the surface gravity is below the Eddington limit) for typical
parameters of AGN disks. We then calculate a model with the
lowest surface gravity for which an atmosphere is stable. Finally, 
for completeness, we will 
compare the predicted flux with a black-body distribution corresponding
to the same effective temperature. 

The classical NLTE stellar atmosphere models are calculated with the program
TLUSTY, using the exactly same model atoms and the same opacities as for the 
disk models. 
Also, the line transitions are considered in the detailed radiative balance,
in order to minimize computational differences between the disk and the
atmospheric
models. The differences in predicted spectra will thus reflect real differences 
between the physical structure of a stellar atmosphere and a disk ring. 
These follow from the three basic features:
i) unlike a stellar atmosphere, the total radiation flux is not constant;
ii) the gravity acceleration is not constant over vertical distance, 
but it varies according to Eq. (\ref{he}); 
iii) the total optical depth of a disk ring is not infinite.

Figure 1 displays the predicted flux for these five
models. There are
two striking differences between the LTE and NLTE/C disk-ring results:
i) a largely diminished Lyman discontinuity, more so in the hotter model; and
ii) the He II discontinuity appears strongly in emission for NLTE models.
Consequently, the number of ionizing photons in the He II Lyman continuum
increases significantly for NLTE disk models, which may have profound
consequences for models of intergalactic matter. This feature was first
noted by Coleman (1994) and Shields \& Coleman (1994),
who have however used a very approximate models.
The ``partial NLTE'' model provides a reasonably good approximation to 
the consistent NLTE model; 
the predicted EUV flux is systematically higher by 10 - 20\%.

\begin{figure}
\epsscale{0.8}
\plotone{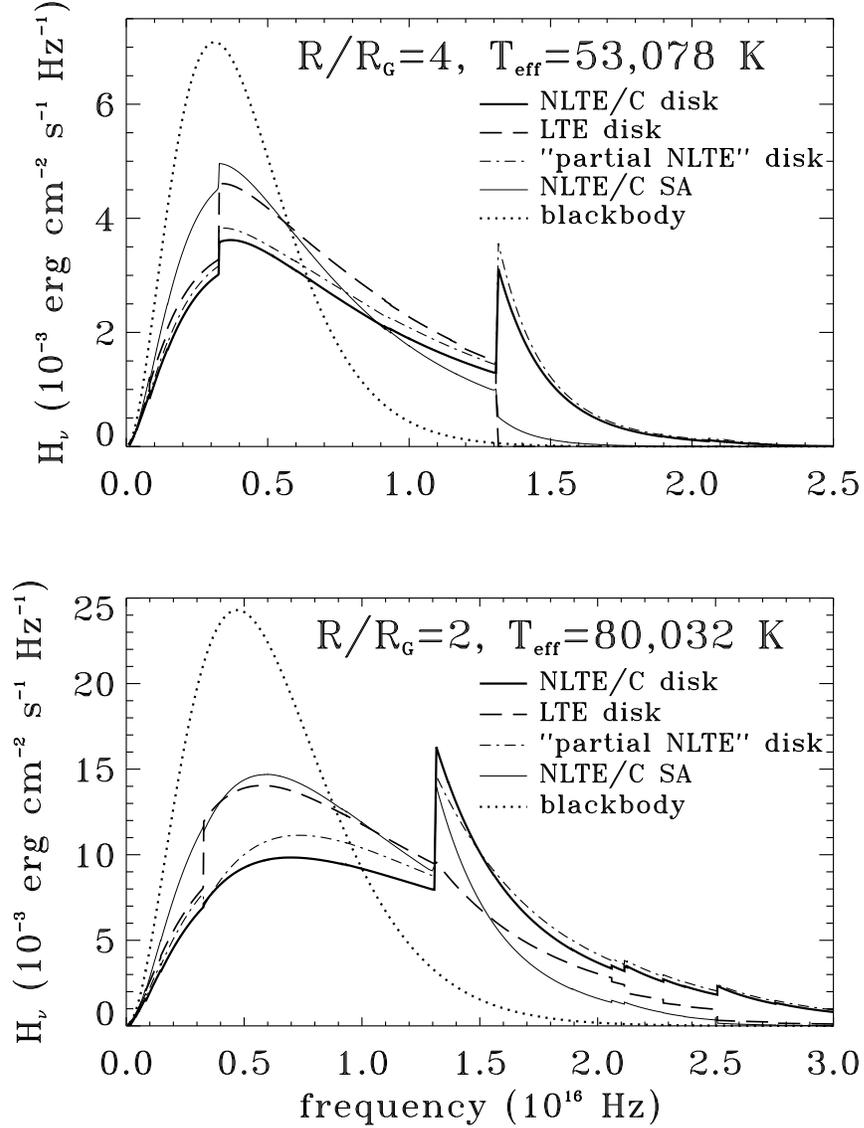}
%\plotfiddle{fig1.eps}{5.5in}{0}{100}{100}{0}{0}
\caption{
A comparison of the predicted spectra for an AGN disk model with a
black hole mass, $M = 2 \times 10^9 M_{\odot}$, a limiting stable rotation
(the specific angular momentum $a/M=0.998$), and an accretion rate,
$\dot{M}=1~ M_\odot{\rm yr}^{-1}$, for two radial distances 
$R/R_{\rm g} = 2$ (lower panel) and 4 (upper panel). 
The corresponding effective temperature $T_{\rm eff}$ is given in
each panel. Five different
predicted spectra are displayed for each ring -- see the text for explanation.
The displayed model stellar atmospheres were computed for the indicated 
effective temperature, and for $\log g$ = 3.8 (upper panel),
and 4.5 (lower panel), which represent the lowest gravities for which the 
respective hydrostatic model atmospheres can be constructed.
}
\end{figure}

A comparison between the NLTE/C stellar atmosphere and the NLTE/C disk models 
reveals
significant differences: The flux in the He II Lyman continuum is systematically
lower, while the flux in the H Lyman limit is systematically higher, in the
stellar atmosphere models. 
Finally, the black-body approximation is very inaccurate, 
giving much too high a 
flux in the UV region ($\nu \approx 3\, -\, 4 \times 10^{15}$ Hz), while
too low a flux in the EUV and the soft X-ray region.
In particular, the number of ionizing photons in the He II Lyman continuum
for the black-body model is by factor 15 and 75 lower than that for the NLTE
disk model, for the 
$R/R_{\rm g} = 2$ and 4 models, respectively!

Since several studies in the past used a stellar atmosphere
approximation,
we plot in Fig.~2 the predicted flux for NLTE/C model atmospheres
with $T_{\rm eff}$ = 53,078 K (corresponding to the 
ring at $R/R_{\rm g} = 4$), and with various gravities.
Since the presence of metals (CNO) does not change
the emergent spectrum significantly, we consider H-He models for 
simplicity.
The Lyman jump goes from a strong absorption for
$\log g > 3.9$, to the emission at $\log g$ = 3.8. At $\log g$ = 3.85,
the discontinuity practically disappears. 

\begin{figure}
\plotone{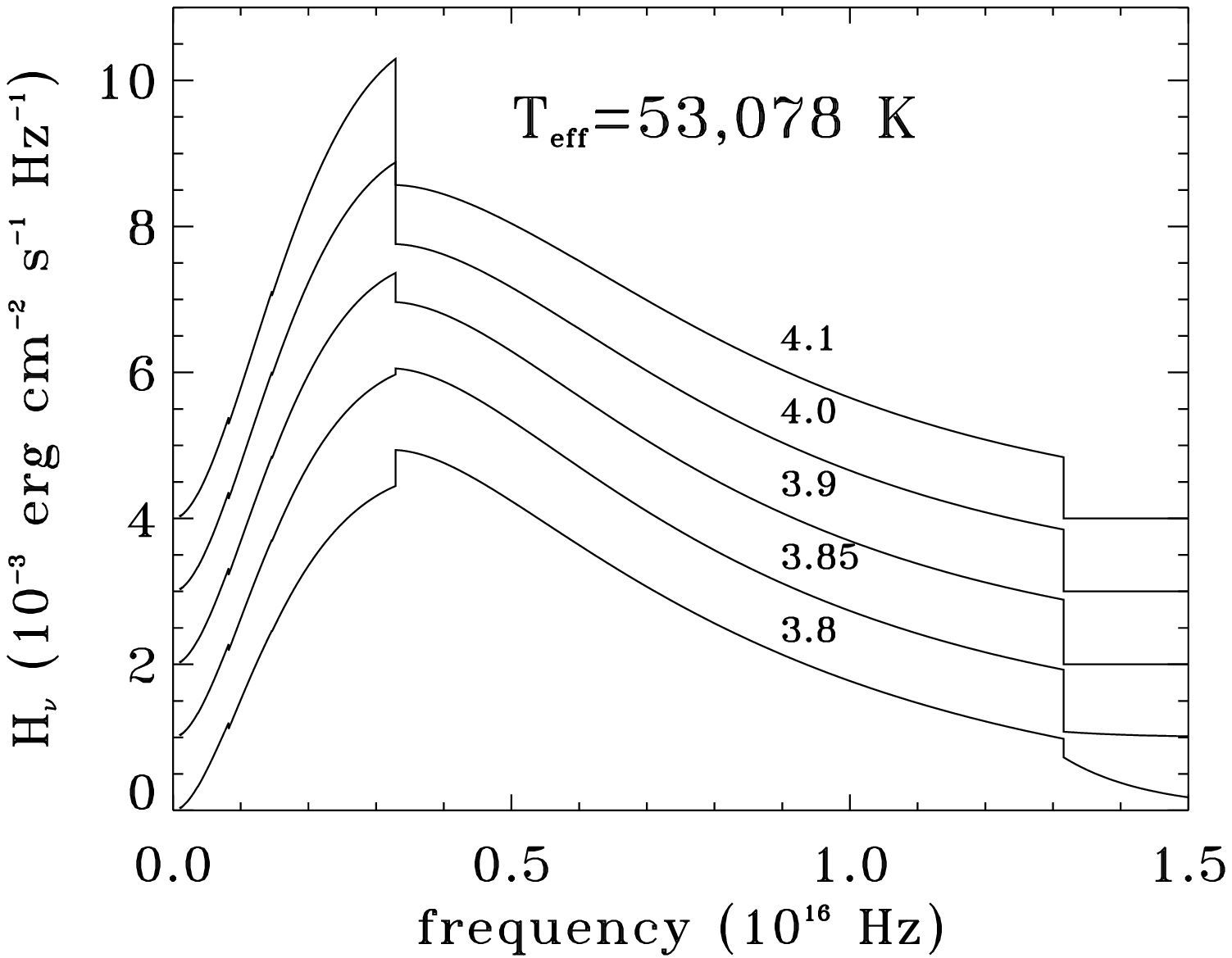}
\caption{
A behavior of the H I and He II Lyman discontinuity for various
NLTE/C H-He model stellar atmospheres with $T_{\rm eff}$ = 53,078~K, and  
for various values the surface gravity. The curves are labeled by the
values of $\log g$. The scale on the ordinate applies for the $\log g = 3.8$
model; the subsequent models are offset each by 1 unit for a clearer
display.
}
\end{figure}

The behavior is easily understood by invoking the Eddington-Barbier relation,
$F_\nu(\tau=0) \approx  S_\nu (\tau_\nu = 2/3)$. 
For all models, the hydrogen ground state is 
underpopulated, while the $n=2$ state is overpopulated in the region
of formation of the Lyman continuum. The thermal source function blueward of
the discontinuity is roughly given by 
$S^{\rm th}(\nu > \nu_{\rm L}) \approx B_{\nu}(T_{\rm b})/b_1$, 
$b_1$ being the departure coefficient for the $n=1$ level, while at the redward
side it is roughly given by 
$S^{\rm th}(\nu < \nu_{\rm L}) \approx B_{\nu}(T_{\rm r})/b_2$. 
Here, $T_{\rm b}$ and $T_{\rm r}$ are the local temperatures at the depth where the
monochromatic optical depth just blueward and redward of the Lyman
discontinuity, respectively, is equal to 2/3. Since the opacity blueward
of the jump is larger, the depth of $\tau_\nu \approx 2/3$ is located higher up
in the atmosphere where the local temperature is lower. Consequently,
$T_{\rm b} < T_{\rm r}$. This leads to the presence of absorption jump in LTE,
where $b_1=b_2=1$.
In NLTE, the magnitude of the Lyman jump is reduced, or it may even 
be reversed.
When going to lower gravities, the portion of electron scattering
increases, which finally becomes the dominant opacity source throughout 
the atmosphere.
Both sides of the Lyman discontinuity are now formed in almost the same
region, so that $T_{\rm b} \approx T_{\rm r}$. 
Since $b_1<b_2$, the thermal source function 
blueward of discontinuity is larger than that redward of the jump.
Moreover, the portion of electron scattering is lower for 
$\nu \geq \nu_{\rm L}$,
which further increases the total source function blueward of the jump.
Consequently, the jump appears in emission.
The explanation of the He II Lyman jump is similar.
The behavior of the Lyman jump for a disk ring model is analogous;
the effect is larger because the effective gravity 
is even lower than in the case of a classical stellar atmosphere.

%
% ====================================================
%

\section{CONCLUSIONS}

We have calculated several representative models of vertical structure
of an accretion disk around a supermassive Kerr black hole. The interaction
of radiation and matter is treated self-consistently, taking into account
departures from LTE for calculating both the disk structure and the
radiation field. 

We have demonstrated that NLTE effects, as well as the effects of 
self-consistent vertical structure of a disk, play a very important role 
in determining the emergent radiation. 
We have shown that NLTE models exhibit a largely diminished H I Lyman 
discontinuity when compared to LTE disk models, in agreement with the
results of St\"orzer, Hauschildt, \& Allard (1994), and Shields \& Coleman
(1994).  The most interesting
result is that the He II discontinuity appears strongly in emission for 
NLTE models, which confirms previous exploratory calculations by 
Shields \& Coleman (1994).
Consequently, the number of ionizing photons in the He~II Lyman continuum
increases significantly for NLTE disk models. 
For the two representative models considered here, the number of He II ionizing
photons predicted by NLTE disk models is larger by factor 15 and 75 than 
that following from the black-body approximation.
This feature may have very important
implications for ionization models of the AGN broad line region
(e.g. Baldwin et al. 1995), and for
models of intergalactic radiation field
and the ionization of helium in the intergalactic medium. 

We have also compared the models of a disk ring with the 
NLTE stellar atmosphere models computed for the same effective 
temperature.
We conclude that using stellar atmosphere models for
approximating the AGN disk emergent radiation is not recommended because 
the predicted flux in the vicinity of the H I and He II Lyman discontinuities
is very sensitive to the assumed value of the surface gravity.
Moreover, even the ``best'' stellar atmosphere model, i.e. the one with the 
lowest surface gravity possible, yields too low a flux
in the He II Lyman continuum, and too high a flux in the H Lyman continuum.

Our results show that in order to be able to test observationally 
the AGN accretion disk paradigm, it is necessary to
construct sophisticated non-LTE models of the vertical disk structure, taking
into account all relevant opacity sources. It is dangerous to use
simplified models, based on LTE approximation, on a vertically constant 
gravity acceleration, or on approximating the
disk radiation by that of a stellar atmosphere or even a black-body. 
Such simplified models may
lead to incorrect predicted spectrum and thus to erroneous conclusions about 
the existence and properties of accretion disks in active galactic nuclei.

%
% ====================================================
%

\acknowledgments
This work was supported in part by NASA grant NAGW-3834 to I.H., and by 
a summer student traineeship to V.H. made possible by the Space Telescope
Imaging Spectrograph (STIS) science team.
We also thank Sally Heap and Eric Agol for helpful comments to the manuscript,
and Matt Malkan and Omer Blaes for valuable discussions.

%
% ====================================================
%

\end{document}